\documentclass[referee]{aa} 
\usepackage{graphics}

\begin{document}

 %

   \title{On the band-to-continuum intensity ratio in the infrared spectra of interstellar carbonaceous dust}
\author{R. Papoular
\inst{1}
          }

   \offprints{R. Papoular}

   \institute{Service d'Astrophysique, CEA Saclay, 91191 Gif-s-Yvette, France\\
              e-mail: papoular@wanadoo.fr
             }

   \date{ }
   \authorrunning {R. Papoular}
   \titlerunning {Infrared band-to-continuum ratio}
   \maketitle
\begin{abstract}
Published interpretations of the relative  intensity variations of the Unidentified Infrared Bands (UIBs) and their underlying continuum are discussed. An alternative model is proposed, in which a single carrier for both emits a) mostly a continuum when it is electronically excited by photons (visible or UV), or b) exclusively the UIBs, when only chemical energy is deposited by H capture on its surface, inducing only nuclear vibrations. The bands will dominate in atomic H regions but will be overcome by thermal continuum radiation when the ambient field is strong but lacks dissociating photons (900-1100 $\AA$). The model applies to PDRs as well as to limbs of molecular clouds in the ISM and agrees quantitatively with recent satellite observations. It gives indications on atomic H density and UIB intensity provided the ambient radiation field is known. It invokes no chemical, electronic, structural or size change in order to interpret the observed intensity variations.

 \keywords{ISM: dust, infrared:ISM, atomic processes}
 %
  \end{abstract}

%
\section{Introduction}

The increasing number and sophistication of satellite and ground-based infrared observations have provided a detailed knowledge of the UIBs (Unidentified Infrared Bands) and their underlying MIR (mid-infrared) continuum in various environments. These measurements cover a range of many orders of magnitude in ambient radiation field intensity and a wide range in incident photon energy, thus setting strong constraints on dust models and physical processes in action.

The observations show that 
\begin{itemize}
\item
the contrast (ratio of band to continuum intensities) covers a very wide range from much less than one to much higher than one;
\item
it is highest in emission in the general, or diffuse interstellar medium (DISM; see Kahanp$\ddot{\rm a}\ddot{\rm a}$ et al. \cite{kah}), far from all illuminating stars; also in Photodissociation Regions (PDRs), at the interface between HII regions and molecular clouds; 
\item
it is low in emission in novae and in some planetary nebulae, and very low in absorption towards protostars and the GC (Galactic Center);
\item
the width and relative intensities of the UIBs hardly change with the contrast.
\end{itemize}

The following sections are devoted to the discussion of published interpretations of these findings, and to the presentation of an alternative model. The latter is constrained by, and confronted with, available observations.

\section{The 3-component model}
The variations of contrast are usually discussed in terms of the 3-component model. As first proposed by Desert et al. \cite{des}, this model invokes PAHs (Polycyclic Aromatic Hydrogenated free-flying molecules), VSGs (Very Small Grains, assumed to have solid-state properties) and BGs (Big Grains in thermal equilibrium) as being respectively the \emph{exclusive} carriers of the UIBs, the MIR (Mid-InfraRed) and the FIR (Far IR) continuum. The size limits separating these 3 populations are not clear-cut: PAHs are of molecular sizes, VSGs are nanometric and BGs are larger still. The emissivity of each component is not that of any known material but is tailored once and for all so as to allow the fitting of the different celestial spectra by simply adjusting the relative concentrations of the three components. Thus, each of the latter can be likened to the empirical ``astronomical silicate" defined by Draine and coworkers (see Li and Draine \cite{li}) on the basis of various astronomical observations, \emph{without specifying the dust composition}. This procedure has been extended by Li and Draine to the 3-component model, raising it to a high degree of mathematical sophistication, which provides good fits to the observations of UIBs.

Although the dissociation of the 3 spectral contributions and carrier functions considerably eases the task of numerical fitting, it raises several qualitative and quantitative issues. First and foremost is the need to justify the separate adjustment of the 3 dust concentrations in each astronomical environment, a procedure which is not deemed necessary for silicious dust, although large contrast variations are also observed in the latter case. The rationale for this has been sought in the more or less complete destuction and/or ionization and/or aromatization and/or simple dehydrogenation by energetic UV photons (see Verstraete et al. \cite{vers}, Vermeij et al. \cite{verm}). However, these processes cannot be adequately quantified until the exact nature of the 3 components is known. As a consequence, discussions of contrast variations are usually limited to a qualitative invocation of the said processes in the conclusions of papers. 

Moreover, the slack definition of compositions and processes does not make for robust interpretations of observations, and so may lead to ambiguous,  and even conflicting, conclusions. This is especially to be expected when energetic photons are freely invoked for carrier processing, or even destruction,  as well as excitation (see below and, e.g., Haas et al. \cite{haa}, Abergel et al. \cite{aber}).

Perhaps the most puzzling observation in this context is the tight correlation observed between the spatial distribution of UIB emission in the DISM and the general distribution of large dust grains (Haas et al. \cite{haa} : sub-mm emission; Kahanp$\ddot{\rm a}\ddot{\rm a}$ et al. \cite{kah}: IRAS 100$\mu$m) and molecular gas (Kahanp$\ddot{\rm a}\ddot{\rm a}$ et al. \cite{kah} : CO). This, for the authors, implies a close relation between the physical processes that are responsible for creating and heating the various component populations. How can such a coherence be achieved with 3 independent components ? A close correlation is also observed between the 4-$\mu$m excess continuum and UIB emission, suggesting they are produced by the same material (Lu et al. \cite{lu}). 

The puzzles relating to the continuum can hardly be dissociated from the other riddles posed by this model (see Sellgren \cite{sel01}). Their solution therefore requires an analysis of its foundations. The Desert model is entirely built upon the twin highly constraining assumptions made 20 years ago: a) the carriers of the UIB and the near- and mid-IR continua must all be very small; b) their emission mechanism is ``stochastic heating", whereby absorption of a VUV photon (10 to 20 eV) by the dust particle is immediately followed by  complete thermalization to a temperature which may (briefly) exceed 1000 K if the grain is small enough.

The first assumption arose from the initial belief that the UIB carriers were one or another of the lightest PAHs (e.g. coronene). Being molecules, these cannot carry any continuum; hence the recourse to VSGs. The second assumption followed from the early finding that the general slope and the relative strengths of bands in the mid-IR spectrum of RNe were nearly independent of distance from the star (Sellgren \cite{sel84}). Single-photon ``heating" was adopted as an explanation, because, in that case, the spectral energy distribution depends on the photon energy and grain size, not on the radiation flux density. The same argument is invoked nowadays to explain the stability of band intensity ratios over orders of magnitude variations of radiation fluxes (see Boulanger et al. \cite{boul}, Kahanp$\ddot{\rm a}\ddot{\rm a}$ et al. \cite{kah}). 

Later observational and theoretical results seem to question the validity of, and the need for, these assumptions, as we now set out to show.

\section{Small sizes}
The first doubts about the smallness of PAHs were raised by Sellgren et al. (\cite{sel90b}b) who used IRAS spectra of RNe to show that the ratio of energy radiated in the 12 $\mu$m band to total energy is independent of the effective temperature of the illuminating star over the range 5000-22000 K. This fact requires that the material responsible for the UIBs (which dominate the 12-$\mu$m signal) absorbs not only in the UV (which is not emitted by the coolest stars) but also at visible wavelengths. But this is not possible for small particles because the absorption cut-off recedes to shorter wavelengths as the particle size decreases. By the same token, if VSG's are as small as required for emission, then they can hardly carry the continuum they are supposed to provide.

No less compelling is the fact that, despite the efforts of many laboratories over two decades (see Boulanger et al. \cite{boul}, Hony et al. \cite{hon}, Peeters et al. \cite{pee}, Sellgren \cite{sel01}), no combination of \emph{molecular-size} PAHs has been shown to mimick convincingly the observed UIBs in position, bandwidth and relative intensities.

Moreover, the 3-component model implies that the energetic photons will heat the smallest PAHs to temperatures of 1000 to 3000K. Now, in  emission, highly excited molecules may have much broader features than in absorption because of \emph{anharmonicity and spectral congestion}, as observed by Williams and Leone (\cite{wil}) on naphtalene molecules excited by short laser pulses of photons of energy 5.3 and 6.6 eV. Monitoring the outgoing radiation at 3.3 $\mu$m, (C-H stretch), they found its width to decrease from $\sim$100 to $\sim$30 cm$^{-1}$, while the peak shifted to the blue by 45 cm$^{-1}$, as the energy content of the molecule was reduced by on-going radiative emission. Boulanger et al. (\cite{boul}) noted that no such behaviour is observed in the sky.

 Faced with these difficulties, many authors now seem to have condiderably relaxed the size constraints (see Boulanger et al. \cite{boul}), so the terms ``VSG", and even ``molecular carriers" (subsumed in the word PAH) may be misleading. More importantly this evolution of the model leads to two inconsistencies:

\begin{enumerate}

\item
If the PAH size is increased enough for it to absorb enough visible light so as to be able to emit bands similar to the UIBs even under low effective temperature irradiation, then, it is no longer possible to ignore its own MIR continuum, which is incompatible with the separation of functions of the three components. 
\item
If the PAHs are big enough, and if stochastic heating is retained as the excitation process, then, available photon energies (limited to 13.6 eV by atomic hydrogen ionization) cannot raise them to the required temperature for UIB emission, which results in low contrasts.

\end{enumerate}

Let us consider the first issue somewhat more quantitavely. PAHs are made up of coalesced benzenic (aromatic) 6-membered plane carbon rings. As the $\pi$ electrons which contribute to the carbon bonding are relatively free to move around the rings, conductivity is to be expected in the plane of the latter. A free-electron model, proposed long ago by Platt, indicated a conduction band with a lower energy cut-off inversely proportional to the aromatic surface area. More recently, Robertson (\cite{rob}) developed a more realistic model which, starting from an isolated C-C bond, and going through compact or linear cluster rings, to graphitic layers, clearly shows the transition from non-conducting molecules with only sparse, narrow, discrete electronic energy levels and, therefore, devoid of continuum absorption, to \emph{semi-conducting} solids: as the number M of clustered rings increases, the density of attainable energy levels increases rapidly to form a conduction band. While the upper energy cut-off of the latter falls in the VUV ($\sim$10 eV) and does not vary much, the lower energy cut-off (the semi-conducting band-gap energy) decreases steadily to zero. For compact clusters, this scales as M$^{-0.5}$. Experimental results published much earlier by Inokuchi et al. (\cite{ino}) indicate a similar but much faster trend: 1.6 eV for 20 carbon atoms down to 0.8 eV for 35 atoms.

Now, the structure of dust is expected to be disordered like that of a-C:H, for instance. Then, the steep cut-off  of regular structures is replaced by an extended edge (or tail). As a consequence, the conductivity, and, hence, the continuum absorptivity, $\alpha$, fall off slowly into the IR (see Rouleau and Martin \cite{rou}): for glassy or amorphous carbons and highly aromatic coals, $\alpha$ is still of order 10$^{4}$ cm$^{-1}$ near $\lambda$=10$\mu$m.

For kerogens and less evolved (aromatic) coals, the MIR continuum is much weaker but still far from negligible with respect to the band intensities (see Guillois \cite{gui}). Thus, even in the absence of VSGs and BGs, no known material, \emph{if assumed in thermal equilibrium with the exciting radiation flux}, could produce such a high UIB contrast in emission  as observed in the DISM (see Kahanp$\ddot{\rm a}\ddot{\rm a}$ et al. \cite{kah}).

Independent of any excitation process, \emph{absorption} spectra unambiguously provide the relative values of continuum and band absorptivity (or imaginary index of refraction). However, in the sky, UIBs are hardly observable in absorption: only the 3.4-$\mu$m band has clearly been identified, and then only as due to various aliphatic carbon bonds, and only towards very dense regions, such as the GC (Galactic Center; see Chiar et al. \cite{chi}) and protostars (see Bregman et al. \cite{breg}). This might be ascribed to a high abundance of an unspecified continuum carrier (not necessarily in very small sizes like VSGs). But that would  not improve our knowledge of IS dust. Besides, it seems difficult to reconcile very high abundances of continuum carriers towards the GC and protostars with the total absence of the same carriers in the DISM. Moreover, the continuum that was shown above to be provided by solid-state carbon grains is better suited to this function since it is known to be sensitive to ageing, heat and radiation, so the mid-IR contrast it provides can be tailored to a large extent so as to fit observations (Papoular et al. \cite{pap96}, Papoular \cite{pap01a}). Further, this continuum extends into the near-IR and beyond, as required by the galactic extinction curve and the near-IR excess emission.

Thus, it would seem that, from the spectral point of view, dilemma (1) above can be solved by doing without separate PAHs and VSGs, using instead only partly aromatic natural materials like kerogens (see Papoular \cite{pap01a}). But, with grains of ``normal" sizes, we run into dilemma (2), which is addressed next.

\section{Stochastic heating by photons}

On top of the difficulty mentioned in (2) (Sec. 3), this excitation model faces other serious problems. One is that, if the highest energy photons are the \emph{direct} exciting agent, then ionization of the molecule is likely to occur (especially when one trespasses the H-ionization limit of 13.6 eV !). In that case, the characteristic IR spectrum notably differs from that of the corresponding neutral molecule (see Langhoff \cite{lang}). So much so that this sensitivity to ionization was often invoked as a means to allow the fitting of observed spectra by finely tuned mixtures of neutral and ionized PAHs (see Allamandola et al. \cite{all}; Verstraete et al. \cite{vers}). However, this must rely upon the strong but unsupported assumption that the spectral positions and widths of the features are the same functions of temperature for all interstellar PAH sizes, or on very specific mixtures of selected PAH neutrals and cations, supposed to be independent of history and environment (see discussions by Verstraete et al. \cite{vers} and Boulanger et al. \cite{boul}).

Even if ionization does not occur, the model predicts that the dust grain will be raised to a temperature which depends on the ``average" photon energy, i.e. on the hardness of the ambient radiation field and, therefore, on the effective temperature, T$_{eff}$, of the illuminating star. Given the large range of the latter covered by Sellgren et al.'s observations (\cite{sel90b}b), one should expect a sharp decrease of grain temperature, and hence a dramatic change of the emitted spectral distribution. However, the relative UIB intensities from reflection nebulae are observed to be independent of the temperature of the illuminating star from 22000 down to about 5000 K (see Sellgren \cite{sel01}).

Further, observations of several galaxies, more or less rich in star forming regions, provide evidence that no ``extraordinary " conditions (strong VUV flux) are required for UIB emission; only a minimum radiation field, such as that of the DISM (see Haas \cite{haa}). 

This discussion calls into question the role of UV photons and the very notion of ``temperature".

\section{An alternative: chemiluminescence}

Photons act directly on the material's electrons. In a semiconductor, if the photon energy is higher than the band gap energy, the electron will reach the conduction band and excite a continuum emission upon relaxation. If not, only molecular (nuclear) vibrations will be induced, followed by IR band emission with high contrast, as required in the DISM, for instance. However, solid-state aromatic carbons have very small gaps, if any. So, unless the dust particle is very small, even visible photons will excite a continuum emission in the IR, accompanied, perhaps, by very weak band emission, resulting in a low contrast, incompatible with observations.

Chemiluminescence is proposed here, instead. In this process, excitation energy is provided by hydrogen radicals (atoms as opposed to molecules) recombining on grains which need not be of molecular size. Guillois et al. (\cite{gui98}) and Papoular (\cite{pap02}) discussed this mechanism extensively as a substitute to VUV excitation. It is closely related to the 3-body surface recombination known to be the main mechanism for H$_{2}$ production in the ISM. It was applied to the ISM of the VUV-defficient galaxy M31 by Papoular (\cite{pap00}). Recent observations of molecular cloud limbs and PDRs allow a more detailed treatment, leading to a deeper understanding of the parameters involved in the UIB emission, as we now proceed to show.

\subsection{Atomic regions}
Savage et al. (\cite{sav}) have studied the ratio of molecular ($H_{2}$) to atomic ($H$) hydrogen densities in a large number of molecular clouds with various colour excesses E(B-V) and total gas density ($H_{t}=H+2H_{2}$) up to a column density of 2.5 10$^{21}$ cm$^{-2}$. It was found that the molecular fraction is either very low ($<$0.01) or very high ($>$0.1) according to whether E(B-V) is smaller or larger than 0.1-0.2. This transition corresponds to a total hydrogen column density $N_{cr}= 5-10\,10^{20}$ cm$^{-2}$ and a visible extinction A$_{v}$=0.3-0.6 mag. These observations can be understood in the light of theoretical analyses made by several authors who studied the \emph{equilibrium} abundance of $H_{2}$ under the opposite effects of \emph{dissociation} by the IS VUV radiation flux and \emph{recombination} on the surface of IS grains, taking into account the self-shielding of $H_{2}$ molecules against radiation penetrating radially into the cloud (see Hollenbach et al. \cite{hol}; Glassgold and Langer \cite{gla}; Draine and Bertoldi \cite{dra96}). The emerging qualitative picture in 1-D is that, at the surface of a thick molecular cloud, a thin layer of purely atomic H forms under the dissociating action of that small part of the ambient IS radiation with wavelength between about 900 and 1100 \AA. At a depth defined by the transition total column density which is an increasing function of the incident radiation flux, this radiation is suddenly and sharply attenuated by recombined molecular hydrogen absorption; beyond that, the gas is almost entirely molecular. This behaviour is largely insensitive to total hydrogen density or its distribution, and gas or grain temperature.

A PDR can be considered as a portion of such a cloud limb, usually seen edge-on. In this case, however, the radiation emanating from a near-by luminous young star not only dissociates the hydrogen in the limb, but may also be strong enough to drive a shock wave into the cloud. If the shock velocity $V_{s}$ corresponds to a Mach number $M$, the ratio of densities in and beyond the shock front is $M^{2}$ (see Spitzer \cite{spit}). Compression ratios of 100 and more can thus occur in very thin ridges. Since part, or the whole of, the shock will also be dissociated, very high densities of atomic H are encountered there (up to a few 10$^{4}$ cm$^{3}$).

\subsection{Emission model}
Based on the above, we consider an ideal, 1-D, purely atomic region, with uniform total gas density, $H_{t}=H$, permeated by a uniform radiation field density, $F$. According to Spitzer (\cite{spit}), the total cross-section area of IS grains is about 10$^{-21}$ cm$^{2}$ per H atom. For spherical grains, the corresponding surface area is $\Sigma=4\,10^{-21}$. $\Sigma$ will be assumed to retain this value for all objects considered here. If the average relative velocity of grains and gas atoms is $V$ cm.s$^{-1}$, then the rate of atomic impacts in 1 cm$^{3}$ of IS space is ($H\,V/4)\,\Sigma\,H_{t}$. When an H atom impacts the surface of a grain, it can either \emph{abstract} another H atom, bound to a C atom of the grain, to form  a molecule (according to the accepted scenario of molecular formation in the ISM), or it can be captured by a free (dangling) C bond. Such free bonds are bound to always be present, precisely as a consequence of the former process. Let $\rho$ be the probability that an impact ends up in the capture of the H atom. The potential (chemical) energy carried by the free radical ($\epsilon\approx$4.5 eV, the energy of the C-H bond) is then deposited in the grain in the form of vibrational energy of the grain nuclei, \emph{with the system remaining in the ground electronic state}.

It was argued by Papoular (\cite{pap02}; this is a specialized review but a summary of the argument is given here in Appendix A) that this energy is quickly, and almost equally, partitioned between the vibrational modes of the grain . Because of the strict isolation of grains in space, the only outlet for this energy is radiation by allowed transitions, i.e. those with strong electric dipole moment. These are known to fall almost all in the near and mid-IR, 3 to 15 $\mu$m, i.e. precisely the UIB spectral range. We shall therefore assume that all the free radical H energy is reemitted in the UIBs, with intensities proportional to their respective oscillator strength and illustrated by the spectra gathered from many directions in the Galaxy (e.g. see Kahanp$\ddot{\rm a}\ddot{\rm a}$ et al.(\cite{kah}), which are all found to be very similar in band position, width and relative intensity. This spectral stability and universality is to be expected from the excitation process envisioned here, because it involves a single chemical reaction, independent of the energy of the ambient photons. Spectral variations, if any, can only be ascribed to chemical changes in the dust.

The total UIB luminescence power emitted by 1 cm$^{3}$ of space in all directions can therefore be written
\begin{equation}
$$P_{l}= \frac{1}{4}\, \rho\, V\, \Sigma\, H_{t}\, H\, \epsilon$$
\end{equation}
or, per cm$^{2}$ of grain surface,
\begin{equation}
$$p_{l}=\frac{1}{4}\, \rho\, V\, H\, \epsilon$$ 
\end{equation}
In these equations, we shall put $\rho= 1/3, V=10^{6}$ \,cm.s$^{-1}$, $\Sigma=4\, 10^{-21}$ cm$^{2}$ per H at. and $\epsilon$=7.2\, 10$^{-19}$\, J.

H capture cannot generate continuum emission, for the latter can only result from excitation of electrons from their ground state into the conduction band of the solid. Such excitations follow absorption, by the same grain, of vis/UV photons of the ambient radiation field. In equilibrium between radiation and absorption and reemission by the grain at longer wavelength, the grain reaches a steady thermodynamic state with temperature $T$. A rough estimate is usually obtained by assuming different coefficients for absorption ($\alpha_{v}$ on average in the vis/UV) and emission ($\alpha_{ir}$ in the IR), and equating total absorbed and emitted thermal powers:

\begin{equation}
$$F \alpha_{v} d \frac{\pi d^{2}}{4}= \pi d^{2}\int_{ir}d\,\alpha_{ir} P(T,\lambda) d\lambda$$,
\end{equation}
where d is the grain diameter, $T$ its temperature and $P$, Planck's function. To the same degree of approximation, we take $\alpha_{ir}=\alpha_{o} \lambda_ {o}/\lambda$ where $\alpha_{o}$ is the absorption coefficient at $\lambda_{o}$ ( absorption index $\beta$=-1) and assume that all the thermal emission occurs at the wavelength of the peak of Planck's function, $\lambda=3000/T$ ($\mu$m, °K). The r.h.s. of eq. (3) can thus be rewritten as 

\begin{equation}
$$\pi d^{3} \alpha_{o} \lambda_{o} \sigma T^{5}/3000$$,
\end{equation}
 where $\sigma$ is Stefan's constant 5.7$\,$ 10$^{-12}$ Wcm$^{-2}$K$^{-4}$. We also write $F=10^{-9} G$, where the ambient vis/UV field in the diffuse ISM, $F_{o}$, is taken to be 10$^{-9}$ W.cm$^{-2}$ and $G$ is a multiplying factor. While the grain temperature is determined by the ratio $\alpha_{v}/\alpha_{o}$, the IR thermal emission is explicitely proportional to $\alpha_{o}$, which varies widely between materials. The latter will therefore be left as a free parameter of the model, while $\lambda_{o}$ will be set at 10 $\mu$m and $\alpha_{v}$ at 2.5 10$^{5}$ cm$^{-1}$, a common value for black carbon materials (see Rouleau and Martin (\cite{rou})). Then,

\begin{equation}
$$T=77\,(G/\alpha_{o})^{1/5}$$,
\end{equation}
independent of the grain diameter. It is now possible to compute the thermal IR emission per cm$^{2}$ of grain surface as
\begin{equation}
$$p_{th}=d \int_{\Delta}\alpha_{ir}(\lambda) P(T,\lambda)\,d\lambda$$,
\end{equation}
in any given wavelength interval $\Delta$.

We now set out to confront the model with available observations in order to a) determine the unknown parameter $\alpha_{o}$ in (4), b) look for correlations between G and H so as to complete the model and reduce the number of free parameters, and c) detect possible inconsistencies in the model.

\subsection{Observational constraints}
The peripheral dust temperature for clouds in the diffuse ISM ($G=1$) is observed to be about 20 K. Dust heated by near-by stars ($G$=100 to 3000) usually reaches 60 to 100 K. Both these constraints are satisfied by making $\alpha_{o}$=800 cm$^{-1}$ in (4), so that $T=20\, G^{1/5}$.

A relation between $H$ and $G$ can be sought in the work of Boulanger et al. (\cite{bou98}), who give the intensities of the main UIBs as a function of the ambient radiation field for $1<G<10^{5}$. Within the framework of the present model, this can be construed as a nearly linear relation between $H^{2}$ and $G$. For the objects observed are cloud limbs and PDRs mostly seen edge-on. For such objects, $H\simeq H_{t}$, so that the total intensity is obtained from (1) as

\begin{equation}
P_{l}L/2\pi=1.2\,10^{-12}H^{2}L,
\end{equation}
in Wm$^{-2}$sr$^{-1}$, where $L$, in pc, is the depth of the object along the sight line. For each $G$, we identify this with the corresponding sum of the feature intensities given by Boulanger et al. (\cite{bou98}).

Now, Kahanp$\ddot{\rm a}\ddot{\rm a}$ et al. (\cite{kah}), observing a similar set of objects with ISOPHOT-S and associating the corresponding measurements of IRAS in the 100-$\mu$m-band (80-120 $\mu$m), find a nearly linear relation between total UIB intensity and 100-$\mu$m brightness (assumed to be totally thermal). When expressed in W\,m$^{-2}$sr$^{-1}$ and W\,m$^{-2}$sr$^{-1}$$\mu$m$^{-1}$ respectively, their ratio is about 13. Using (5) and (6) with $d=10^{-6}$ cm, we express the thermal brightness as

\begin{equation}
\frac {P_{th}L}{2\pi}=1.5\,10^{-3}P(100\mu m)HL,
\end{equation}

in Wm$^{-2}$sr$^{-1}$$\mu$m$^{-1}$, where the Planck function $P$ depends on $G$ through $T$ (eq. 4). Constraining the ratio of (7) and (8) to be 13, we find $H$, and hence $L$, as functions of  $G$. These quantities are plotted in fig. 1.

\begin{figure}
\resizebox{\hsize}{!}{\includegraphics{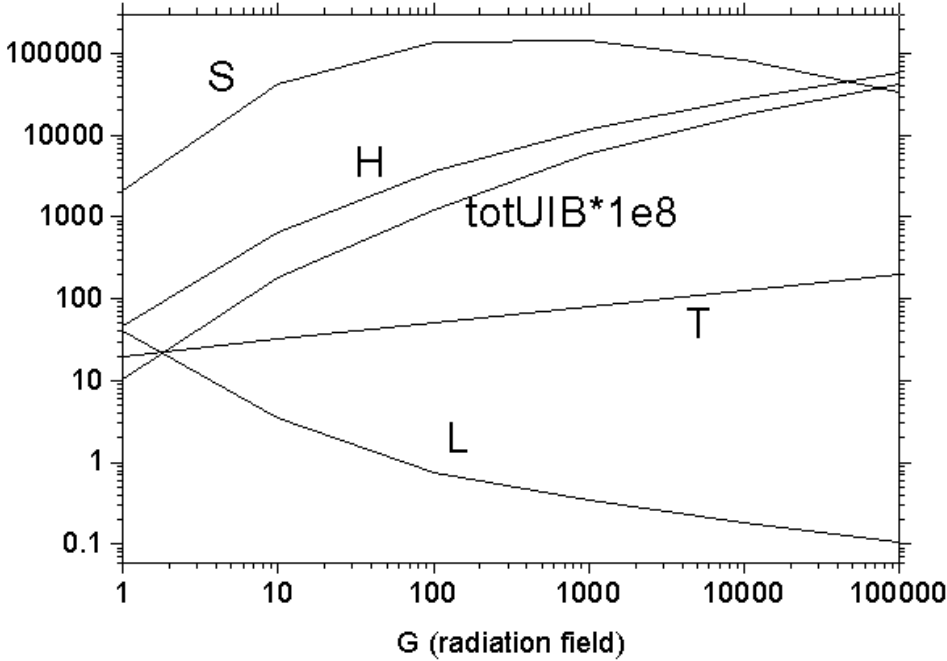}}
\caption[]{ a) TotUIB*10$^{8}$: sum of UIB intensities, in Wm$^{-2}$sr$^{-1}$, estimated from the points in fig.3 of Boulanger; the following are deduced from the model: b) $H$: atomic H in cm$^{-3}$, $L$: depth along the sightline of the observed rim of a PDR, in pc; $S$: shock parameter= $H^{2}/G$,
 in cm$^{-6}$.}
\end{figure}

It should be noted that the range of UIB intensities measured by Kahanp$\ddot{\rm a}\ddot{\rm a}$ et al. is smaller than that of Boulanger et al. \cite{bou98}), so the extrapolation of the linear relation in the former, up to $G=10^{5}$, has to be justified. Indeed, in the present model, this relation imposes that $H$ be proportional to $P$(100 $\mu$m), which in turn induces a relationship between $H$ and $G$ illustrated by the ratio $S(G)\equiv H^{2}/G$ (see fig.1). It is shown in Appendix B that this has a simple physical meaning in terms of gas compression under a shock wave driven by radiation pressure on dust, which drags the dust. Numerically, 
\begin{equation}
$$S=6000\,H_{o}/T_{o}$$,
\end{equation}
where $H_{o}$ and $T_{o}$ are, respectively, the density and temperature of the unshocked gas. Between $G=30$ and $G=3000$, $S$ is roughly constant at 10$^{5}$, corresponding to a reasonable $H_{o}/T_{o}\sim$20. The sharp drop in $S$ at low values of $G$ may indicate weaker shocks or none at all.

At this point, the main physics of the model have been clarified, so that the observables of interest can be deduced from $G$, provided $S$ is chosen correctly. In principle, $S$ could be derived from shock physics. Here, for want of a better knowledge of shock parameters, $S$ was induced from the combined observations of Boulanger et al. (\cite{bou98}) and Kahanp$\ddot{\rm a}\ddot{\rm a}$ et al. (\cite{kah}).

Note that the objects from which fig.1 was deduced are likely to be the brightest observable ones, i.e. PDRs seen edge-on. In such cases, the \emph{depth} $L$ cannot be deduced from first principles, because it is contingent to the particulars of the environment. By contrast, when the line of sight cuts through the star which drives the shock wave, the dimension which determines the relevant column density is the \emph{thickness} of the ridge (or bar), $t$, which is orthogonal to the \emph{depth along the line of sight}, is usually much smaller than $L$, and should then be used, instead, in the equations above. An approximate expression for the column density corresponding to $t$ is given in appendix C.

Using this notion, it is possible to determine an overall efficiency for the chemical excitation process. Consider a star with luminosity $L_{*}$, surrounded by a cloud in which it drives a shockwave. This gives rise to a concentric shell of atomic H of radius $r$ and radial column density $N_{cr}$. Noting that, in the shell, $G=10^{9}\,L_{*}/4\,\pi r^{2}$, one can deduce from eq. 1 the ratio of total band emission to star luminosity

\begin{equation}
$$\eta=2.4\,10^{-4}\frac{H}{G}$\rm ln$(1+36\frac{G}{H})$$
\end{equation}
which is plotted in fig. 2: it spans the range 10$^{-3}$ to 10$^{-2}$. A related ratio, which may sometimes be more easily compared with observations, is $\zeta$, the total UIB emission over total thermal emission of the shell (mostly far IR). This is equal to $p_{l}/p_{th}$ and can be computed using eq. 2 and 6 to give 

\begin{equation}
$$\zeta=\frac{p_{l}}{p_{th}}=4\,10^{3}\,\frac{H}{T^5}$$,
\end{equation}

which is also plotted in fig. 2: it is seen to peak at $\sim10\%$ for low $G$'s and decrease beyond, in qualitative agreement with observations of normal galaxies (see Lu et al. \cite {lu}). Note that the ratio of $\eta$ over $\zeta$ is equal to the fraction of incident radiative energy that is absorbed by grains; it tends to 1 with increasing $G$.

\begin{figure}
\resizebox{\hsize}{!}{\includegraphics{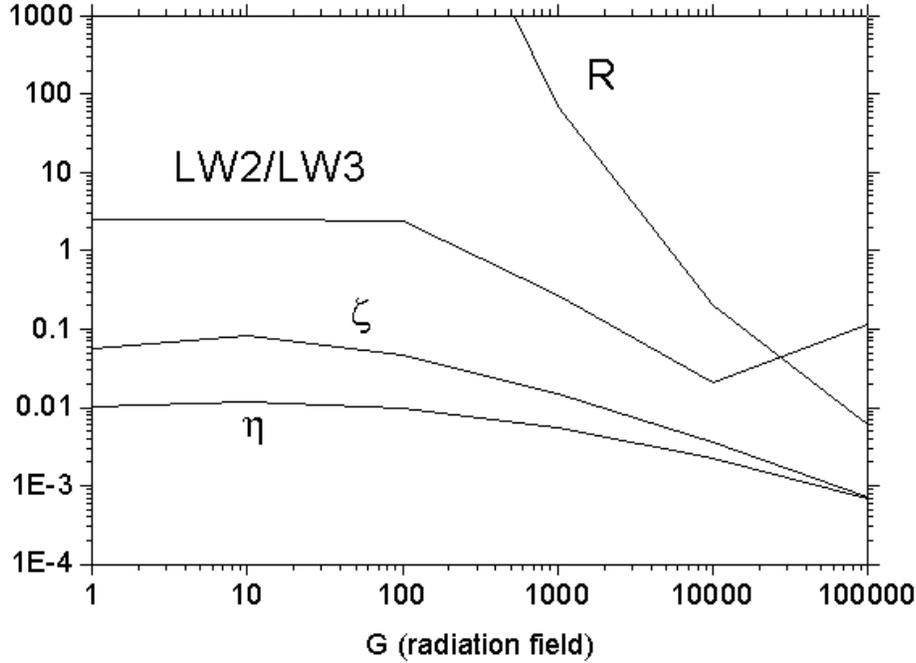}}
\caption[]{Observables deduced from the model: a) $R$, ratio of peak band intensity over thermal continuum at 10 $\mu$m; b) $R_{2/3}$=[LW2]/[LW3], where numerator and denominator are sums of predicted UIB and thermal contributions; c) $\eta$, overall efficiency of chemical excitation, eq. 10; d) $\zeta$, total UIB over total thermal dust emission, eq. 11.}
\end{figure}

\subsection{The band-to-continuum ratio}
Let us now compute the ratio $R$ of the peak UIB brightness to the continuum brightness at 10$\mu$m. Very roughly, the peak UIB intensity is given by the ratio of total UIB power over the sum of the feature widths, $\Delta\sim2.5\,\mu$m. Hence, using eq. (2) and (6),
\begin{equation}
$$R=\frac{\rho\, V\,\epsilon\, H/4\Delta}{d\,\alpha_{ir}(10)\,P(T,10)}$$
\end{equation}
\begin{equation}
$$\;\;\;	=3\,10^{-11}\,H/P(T,10)$$.
\end{equation}

$R$ is plotted in fig. 2.\emph{It is seen to be much higher than 1 in all but extreme cases of strong illumination, in agreement with observations.}

Particular interest also attaches to ISOCAM's bands, LW2 (5-8.5 $\mu$m) and LW3 (12-18 $\mu$m), and IRAS 12-$\mu$m band (8-15 $\mu$m), all of which have been abundantly documented. Using the temperatures deduced for each $G$ in fig. 1, we computed the thermal contributions to these bands, all expressed in W\,m$^{-2}$\,sr$^{-1}$. It was found that the contribution of thermal radiation to LW2 is always much smaller than that of the UIBs. The latter is the fraction of the total UIB spectrum between 5 and 8.5 $\mu$m, i.e. about 50$\%$. Similarly, LW3 includes about 20$\%$ of the total UIB power. In order to compare the model predictions with ISO measurements of [LW2]/[LW3], such as those of Abergel, we added these contributions to the corresponding thermal contributions computed above, and plotted, in fig. 2, the ratio $R_{2/3}$ of the two results. $R_{2/3}$ decreases from a maximum of 2.5 (the ratio of the UIB contributions to the two bands) to very low values as $G$ increases, going through 1 near $G=300$. Here, this trend is interpreted simply as due to the thermal continuum emission increasing more quicly with illumination than does the UIB emission of \emph{the same grain}, so that it overcomes the latter for sufficiently strong illuminations and/or sufficiently low degrees of H dissociation. This may occur, for instance, in novae and planetary nebulae.

On the other hand, when the UIBs are detected in absorption, as is the case towards the Galactic Center, their contrast is the ratio of absorption coefficients of the bulk dust material, in the bands and in the continuum respectively, a very low ratio for all proposed aromatic dust models. This radical difference between absorption and emission contrasts is characteritic of the luminescence model.

\section{Application to Ced 201}
The observation of this UIB emitter by Cesarsky et al. (\cite{ces}) provides a good opportunity to test the present model on a case of spherical geometry with well defined boundaries and a known exciting star. Since the boundaries of the assumed atomic H region move in time under radiation pressure, the characterization of the object at a given moment requires some geometrical information on this boundary to be obtained from observations. In fig. 1 and 3 of Cesarsky et al. (\cite{ces}) the UIB intensity is seen to decrease sharply at a radius of about 11 arcsec (or $r=2.2\,10^{-2}$ pc from center). We shall take this to be the position of the emitting atomic H region. A second quantity is required in order to complete the definition of the object; we take this to be the luminosity of the B9.5V central star, BD+69$\char'27$1231. For such stars, the bolometric magnitude is $\sim$0, so $L_{*}\sim100\,L_\odot$, from which one deduces $G\sim500$ in the atomic region. Our figures 1 and 2 then provide the following model predictions

\begin{equation}
$$T=60\,K,\;H=8\,10^{3} \,\rm{cm}^{-3},\; I_{UIB}=4\,10^{-5}\; \rm{W\,m^{-2}\,sr^{-1}}$$
\end{equation}
\begin{equation}
$$S=10^{5}\, \rm{cm}^{-6},\; R>10^{3},\;R_{2/3}=0.55,\;\eta=6\,10^{3},\;\zeta=0.02$$.
\end{equation}
The model value of $I_{UIB}$ compares well with the value 3.7\,10$^{-5}$, estimated from fig. 3 in Cesarsky et al. by taking a peak brightness of 10 mJy/arcsec$^{2}$ over an effective range of 2.5$\mu$m. From the latter value, and $L_{*}$ and $r$, the ``observed" $\eta=2.1\,10^{-3}$ also compares fairly with the model value, 6\,10$^{-3}$. The predicted very high band-to-continuum ratio,$R$, is also in accord with  the spectra of the same fig.3.

From the observations of Savage et al. (\cite{sav}) and the theoretical results cited above, some degree of hydrogen dissociation is expected to linger beyond $r$. This may explain the low and decreasing  UIB intensities observed in Ced 201 as far as about $2r$. In this range, it seems that the bands are subtended by a weak continuum with an upward slope towards long wavelengths. This is not likely to be a thermal continuum, for, according to Sect. 5.1 and 2 above, the extinction of star light through the atomic region is $\sim1.35$ mag, so that only 1/4 of the star luminosity is available beyond. There, the steady-state temperature of dust must be much lower than 60\,K, which excludes any contribution of thermal emission in the range 5-17\,$\mu$m. Besides, the slope of the weak continuum observed by Cesarsky  et al. (\cite{ces}) is too small to be accounted for by thermal radiation at $T\leq$200\,K. Alternatively, the continuum intensity, $\leq$1 mJy/arcsec$^{2}$, and its slope suggest a zodiacal origin (see fig. 1 of Kahanp$\ddot{\rm a}\ddot{\rm a}$ et al. \cite{kah}).

\section{Conclusion}
The compelling findings of Kahanp$\ddot{\rm a}\ddot{\rm a}$ et al. (\cite{kah}) force us to recognize a strong link between continuum and band carriers, a relationship which transcends all environmental differences, from HII regions to diffuse ISM. In response, the present model takes the stand that they are simply one and the same grain; band or (mainly) continuum emission is their reaction to excitation by atomic H capture or radiative heating, respectively. 

In band emission by H capture, radiation plays only an indirect role: it creates a ubiquitous population of H atoms and, in some regions, enhances band emission by concentrating this population in thin, sheetlike rims. This requires only small amounts of VUV photons and most of the compression is made by visible and UV light, hence the insensitivity to the colour temperature of the ``exciting'' star. The mildness of the direct excitation agent, the H radical, ensures that only the grain nuclei are set in motion, emitting the bands, with the system remaining in the ground electronic state and unable to emit continuum. By the same token, no widening or displacement of bands is expected in this case. Obviously this process is strongly specific of carbon materials in very high vacuum.

On the other hand, with radiative excitation, electrons are set in motion in the first place; they are directly responsible for the continuum but they also induce nuclei vibrations upon their return to their ground state. This produces very weak contrasts, similar to those observed in absorption spectra: they correspond to the low ratio of band and continuum absorptivities characteristic of the carbonaceous materials that are likely to carry the UIBs. Moreover, in emission, the band spectra are expected to depend sensitively on the energy of the exciting photons, a dependence not observed in UIB spectra.

The observable consequences of this model are in approximate quantitative agreement with the observational data considered here. In particular, although radiative and chemical excitations always coexist, the model predicts that, in most situations documented to date, especially in the diffuse ISM, the band-to-continuum ratio will be very high, and fall below 1 only for $G$ larger than about 10$^{4}$. As for the ratio $[LW2]/[LW3]$, the model makes it clear why it should remain around 2 for $G$ much below 100, and decrease through 1 above. Its predictions are also confirmed in the case of the spherical nebulae Ced 201.

In its present state, the model is highly idealized and simplified. Among possible improvements are: more complex geometries and hydrogen density distributions; theoretical treatment of gas and grain density buildup under radiative pressure at the border of molecular clouds; a study of the leeway affordable in the choice of fixed parameters: grain size, absorptivity, etc.

\bf{Appendix A}. \rm IVR vs thermalization

A characteristic of this model is that it does not invoke \emph{immediate thermalization} of the deposited energy. Indeed, the derivation of a (thermal) equilibrium temperature in the stochastic model implies that the thermalization of the absorbed photon energy occurs before radiative relaxation by IR emission. It is true that the latter requires up to 1 s of time, much longer than typical times for transfers of energy from electrons to nuclear vibrations and between the latter (IVR). Unfortunately, common wisdom mixes up IVR (equilibration between vibrational \emph{modes}) with thermalization (equilibration between energy \emph{states}) and the last decades of the past century were necessary to discriminate between various types of \emph{chaos} and between various degrees of \emph{randomization} (see bibliography in Papoular \cite{pap02}). In momentum-configuration phase space, thermal (statistical) equilibrium is the extreme case where all points of the energetically available phase space are equally probable and ultimately visited by the representative point of the molecular system. This stage is only ideal and, in fact, very difficult to reach on a microscopic scale and in the finite time of experiments, but it is easier to approach when dealing with macroscopic assemblies of identical particles interacting with each other and with a ``thermal bath'' (itself an ideal concept !). By contrast, experiments resolved in time, space and wavelength have shown that the vibrations of atoms in an isolated system may retain a high degree of coherence for a long time. \emph{That means that only parts of phase space are visited, that randomization is not complete and that a temperature cannot be defined}. Rather, there is near equipartition between coupled modes all over the spectrum and high frequencies are not particularly inhibited, independent of grain size. I have shown numerically, in a specialized journal, that this is the case for a disordered hydrocarbon system, even as large as 500 atoms (Papoular \cite{pap02}). Experiments such as that of Williams and Leone (\cite{wil}) seem to show that randomization in PAHs is also slow after absorption of a UV photon: time constant longer than 30 $\mu$sec, allowing near-IR \emph{photo}luminescence to be observed before thermalization sets in.

 An extensive survey of the field is available in Les Houches vol. LXXVII: ``Slow relaxations and nonequilibrium dynamics in condensed matter'' (Barrat et al.\cite{bar}). 

\bf{Appendix B}. \rm$H^{2}\propto G$

Consider a grain of radius $a$ in a gas with molecular mass $m$ and number density $H$, illuminated by a radiation flux $G.F_{o}$. If $\alpha$ is the absorption coefficient of the grain material at the radiation wavelength, then the grain is accelerated radially by a force

\begin{equation}
$$F_{rad}=G\,F_{o}\,2\pi a^{2}\,\alpha a/c$$,
\end{equation}
where $c$ is the velocity of light. When the grain reaches velocity $V$ relative to the gas, it is subjected by the latter to an opposing force (see Draine and Salpeter \cite{dra})
\begin{equation}
$$F_{drag}=2\pi\,a^{2}\,kT\,H\,G_{o}(s)$$,
\end{equation}
where $s=(mV^{2}/2kT)^{1/2}$. For $s\geq3$, $G_{o}\sim2\,s^{2}$, which is the square of Mach's number, $M$. Now, the grain reaches its limiting  velocity when the drag force equals the radiative force. On the other hand, in a shock, the compression ratio (shocked over unshocked density) is shown in Spitzer (\cite{spit}) to be
\begin{equation}
$$H/H_{o}=M^{2}$$.
\end{equation}
Combining the last two conditions and assuming the shock is isothermal (same temperature $T$ on both its sides), we obtain $H^{2}=S.G$ with $S$ given by eq. 9.

\bf{Appendix C}. \rm The thickness of the atomic region

Assuming dissociation-recombination equilibrium, the total column density of atomic H along the path of the incident dissociating radiation can be estimated by equating the total number of absorbed photons that gave rise to dissociation, to the total number of recombinations on grains along the same line. The former quantity is $F/m$, where $F$ (cm$^{-2}$s$^{-1}$) is the total incident photon flux and $m$ ($\sim$ 0.15) is the fraction of $F$ that effectively leads to dissociation after absorption. The latter quantity is $RH_{t}Ht=RH_{t}^{2}t$, where $R$ is the surface recombination coefficient  (about 3 10$^{-17}$ cm$^{-3}$s$^{-1}$), and it is assumed that the number density of grains is proportional to the total number density of H nuclei. Hence,

\begin{equation}
$$N_{cr} \sim H_{t}t \sim\frac{F}{mRH_{t}}=\frac{F_{o}}{mR} \frac{G}{H_{t}}$$.
\end{equation}

This, of course, holds only in the linear limit of optically thin absorbing layers. A more complete treatment leads to

\begin{equation}
$$N_{cr}\sim 10^{21} \rm{ln}(1+36\frac{G}{H_{t}})$$.
\end{equation}

 For $\frac{G}{H_{t}}$ between 10$^{-2}$ and 1, this ranges between 3 10$^{20}$ and 3 10$^{21}$. Hence eq. 10.

\end{document}